\documentstyle [12pt,epsf]{article}

\newcommand{\beq}{\begin{equation}}
\newcommand{\dd}{\partial}
\newcommand{\eeq}{\end{equation}}
\newcommand{\bea}{\begin{eqnarray}}
\newcommand{\eea}{\end{eqnarray}}

\newcommand{\f}{\phi}

\begin{document}

\baselineskip 7.5 mm

\def\thefootnote{\fnsymbol{footnote}}

\begin{flushright}
\begin{tabular}{l}
CERN-TH/98-125 \\
astro-ph/9804134
\end{tabular}
\end{flushright}

\vspace{12mm}

\begin{center}

{\Large \bf

Resonant production of gamma rays in jolted cold neutron stars
}
\vspace{18mm}

\setcounter{footnote}{0}

{\large
Alexander Kusenko}\footnote{ email address: Alexander.Kusenko@cern.ch}
\\

\vspace{4mm}
Theory Division, CERN, CH-1211 Geneva 23, Switzerland \\

\vspace{20mm}

{\bf Abstract}
\end{center}

Acoustic shock waves passing through colliding cold neutron stars can
cause repetitive superconducting phase transitions in which the proton
condensate relaxes to its equilibrium value via coherent oscillations. 
As a result, a resonant non-thermal production of gamma rays in the MeV
energy range with power up to $10^{52\pm 1}$ erg/s can take place during
the short period of time {\it before} the  nuclear matter is heated by the
shock waves. 

\vspace{12mm}

\begin{flushleft}
\begin{tabular}{l}
CERN-TH/98-125 \\
April, 1998
\end{tabular}
\end{flushleft}

\vfill

\pagestyle{empty}

\pagebreak

\pagestyle{plain}
\pagenumbering{arabic}
\renewcommand{\thefootnote}{\arabic{footnote}}
\setcounter{footnote}{0}

The recent discovery of the afterglow associated with the gamma ray bursts 
(GRB)~\cite{afterglow} has provided convincing evidence that the GRB
originate at redshift $z\sim 1$ from an event that releases of order 
$10^{52}$ erg in photons (assuming the spherical symmetry of emission).  
A number of theoretical attempts to explain the phenomenon invoked a small
energetic fireball~\cite{fireballs}.   Although many astrophysical events,
for instance, neutron star collisions, can release the required amount of  
energy, it is extremely difficult to reconcile the optical thickness of a
fireball with the short duration and the high energy of the gamma ray burst
it is supposed to produce.  In this Letter we describe a new mechanism for
the gamma-ray production by the colliding cold neutron stars prior to the
eruption of a fireball.   We comment on the possible connection of this
phenomenon with the observed GRB.  

We will show that  powerful acoustic shock waves passing
through a neutron star can cause repetitive superconducting phase
transitions during which the 
relaxation of the order parameter is accompanied by a resonant non-thermal
production of  photons with energies of order MeV and the total power
up to $10^{52 \pm 1}$~erg~s$^{-1}$.  These gamma-rays are copiously
produced due to a coherent motion of the superconducting proton
condensate.   Such oscillations can be powered by the density waves
generated, for example, in the collision of two neutron stars. The
production of gamma rays continues until the medium is heated to the
temperature comparable to $T_c$, the superconducting phase transition
temperature.

When two neutron stars collide, the kinetic energy is transfered into
the shock waves that propagate through the nuclear matter and eventually 
dissipate their energy into heat.  The density variations in a
relativistic shock wave are much slower than the time scale of nuclear
interactions. The interior of the neutron star is a superconductor because
it contains the superfluid proton condensate~\cite{super}.  
An acoustic wave excited by the initial impact of a collision, or by
some precursory tidal effects, causes large variations of density in some
macroscopic domains of the size of order the wavelength.  
The proton energy gap depends on the density~\cite{egap} as shown
in Fig.~1; the shock wave can drive it to zero in some formerly
superconducting regions, or vice versa.  Suppose a surge (or an ebb) of
density destroyed superconductivity in some macroscopic region of the star.
At a later time, as the sound wave moves on, the original density  is
restored and the phase transition to a superconducting phase takes
place. The proton condensate $\phi$ changes from zero to the value
$\phi_0$, which minimizes the free energy. This relaxation process is
described by the generalized, time-dependent Ginzburg-Landau (GL)
theory~\cite{at}. In nuclear matter $T_c \sim 0.5$~MeV. At low
temperature $T\ll T_c$, $\phi$ oscillates until it settles in the minimum
(Fig.~2).    

\begin{figure}
\setlength{\epsfxsize}{3.3in}
\centerline{\epsfbox{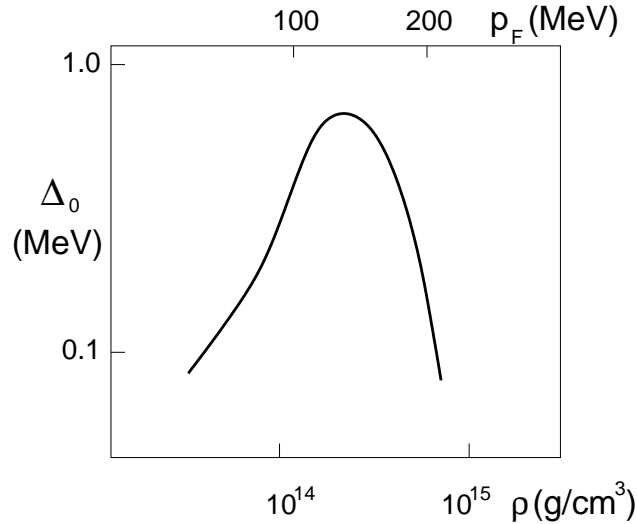}}
\caption{Proton energy gap $\Delta_0$ at zero temperature as a function of
  density $\rho$ and the Fermi momentum $p_{_F}$ of protons; based on
  Ref.~\cite{egap}.  
}
\label{fig1}
\end{figure}

Coherent oscillations of the order parameter result in a non-thermal
production of electromagnetic radiation.   We will see that such emission
is very effective because it occurs through a parametric resonance that
produces some very large occupation numbers of photons in the MeV energy
band.  A resonant emission of Bose particles by a coherently oscillating
scalar field has been studied intensely in quantum field theory, in
particular, in application to cosmology~\cite{cosmo}.  We will 
adopt a very similar approach to a semiclassical description of the
gamma-ray emission via parametric resonance. 

Eventually, the star is heated to the temperature above $T_c$ by the
acoustic shock waves.  At this time the proton condensate is destroyed, the  
resonant emission of photons seizes, and the fireball erupts.  

Now we estimate the energy associated with the coherent motion of the
proton condensate.  As shown in Fig.~1, the energy gap depends sharply on
density.  Let us consider a propagation of an
(ultra-relativistic~\cite{ac}) acoustic shock wave of
sufficiently high amplitude through the interior of the neutron star.  
Density waves are characterized by the time scale
$\tau_a \sim 10^{-7}...10^{-3}$ s~\cite{spectrum}, and are very slow in
comparison to the relaxation time of the proton condensate which is of
order MeV, or $10^{-20}$ s.  The surge of density in the acoustic wave
converts the protons from the superconducting to the normal state. As the
density subsides, the superconducting transition takes place.  The
energy change associated with proton condensation in the volume $V$
is~\cite{ll}

\beq
\delta E  = V \frac{m_* p_{_F}}{4 \pi^2} \Delta_0^2,  
\label{dE}
\eeq
where $m_*$ is the effective proton mass in nuclear matter, $m_*\approx 0.6
m_p$~\cite{egap}; $\Delta_0$ is the proton energy gap; and $p_{_F}$ is the
Fermi momentum. For density $\rho \sim 10^{14} ... 10^{15}$~g~cm$^{-3}$,
$p_{_F} \sim 10^2$~MeV and  $\Delta_0 \stackrel{<}{_{\scriptstyle \sim}}  
0.8$~MeV~\cite{egap}.  This yields $\delta E/V \sim 10^3 ({\rm  MeV})^4$.  

The proton condensate must dissipate energy $\delta E$ in each cycle of the
acoustic ``pumping'' that repeat after time $\tau_a$.  Therefore, the power
dissipated by the condensate in volume $V\sim (4\pi/3) R^3$, where
$R\approx 10$~km is the radius of a neutron star, is 
\beq
P = \frac{\delta E}{\tau_a} \approx \left ( \frac{0.1 \mu {\rm s}}{\tau_a}
\right) \, 10^{52} \ {\rm erg \ s^{-1}}.
\label{power}
\eeq
Analysis of the oscillation spectra of the neutron stars~\cite{spectrum}
gives the range of values for the periods of various eigenmodes:  

\beq
10^{-2} \mu {\rm s} \stackrel{<}{_{\scriptstyle \sim}}
\tau_a 
\stackrel{<}{_{\scriptstyle \sim}}
0.1 {\rm s}
\label{taurange}
\eeq
for magnetic field $\sim 10^{12}$ G.  A larger magnetic field would give
rise to modes with shorter periods $\tau_a$. Clearly, the higher-frequency
modes provide more frequent pumping, forcing the condensate to dissipate
more energy.  

If a sizeable fraction of the energy (\ref{power}) stored in the motion of
the proton condensate at the onset of a phase transition is radiated away
with gamma rays, the power of the resulting signal is consistent with that 
of observed GRB for the range of $\tau_a$ given by (\ref{taurange}).  We
will now describe the resonant emission of photons by the condensate.

\begin{figure}
\setlength{\epsfxsize}{3.3in}
\centerline{\epsfbox{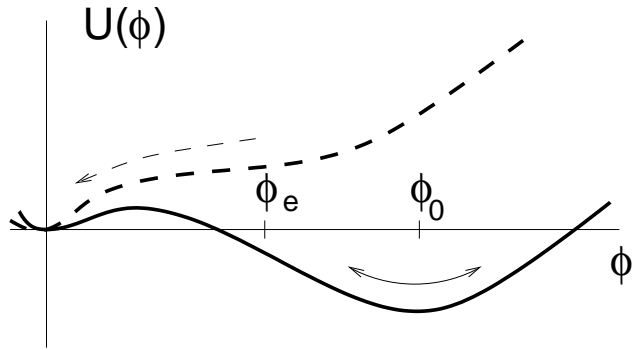}}
\caption{After the phase transition, the order parameter oscillates around
  the minimum of the effective potential.
}
\label{fig2}
\end{figure}

Superconducting condensed-matter systems out of equilibrium display a
number of interesting phenomena related to time dependence of the order
parameter.  However, in the case of a neutron star, the
specific relative values of parameters, the MeV energy scales, 
and the sharp dependence of the energy gap on density make it difficult to
find a close condensed-matter analogue.   Nevertheless, one can use the
general formalism developed for nonequilibrium superconductivity.  

In the presence of a magnetic field, the phase transition is first
order~\cite{super}; it proceeds through nucleation of bubbles that grow and
coalesce.  This process is described by a time and coordinate dependent GL 
order parameter $\phi(x,t)$~\cite{at}.  For simplicity, we will neglect
the effects of magnetic field on the motion of the proton condensate after
the nucleation of the initial bubbles, whose $t=0$ profile, $\phi(x,0)$, 
serves as the initial condition for the equation of motion (\ref{du})
written below.  This is 
a good approximation as long as the magnetic field $H$ is well below the
critical value $H_c$.  If the potential barrier (Fig.~2) is much smaller
than $\delta E/V$ in equation (\ref{dE}) (which again requires $H \ll
H_c$), the maximal value of the field inside the critical bubble, $\phi_e
\equiv \max_x  \phi(x,0)$ is not close to $\phi_0$.  The subsequent 
relaxation of the order parameter $\phi$ (assumed to be real) 
is described by a partial differential equation~\cite{at} 

\beq
\ddot \phi + \frac{8 \varepsilon_{_F}}{3 c} \dot \phi 
- \frac{2\varepsilon_{_F}}{3 c m_*} \nabla^2 \phi
+\frac{\dd U(\phi)}{\dd \phi}
= 0.  
\label{du}
\eeq
Here $\phi(x,t)=(n_p/2m_*)^{1/2} \left ( \frac{\Delta(x,t)}{\Delta_0}
\right )$ is the order
parameter proportional to the energy gap $\Delta(x,t)$; $\Delta_0$ is the
equilibrium value (Fig.~1) that corresponds to
$\phi_0=(n_p/2m_*)^{1/2}\approx 20$~MeV. (Note that we have included the
proton mass $m_*$ in  the normalization of $\phi$, so as to make its
dimension one in mass units; the GL order parameter is often normalized
differently~\cite{at,ll}, in which case $m_*$ reappears in the gauge
coupling.) $\varepsilon_{_F}$  is the Fermi energy; $c=(28 \zeta(3)/3
\pi^3) \varepsilon_{_F}/T_c$; $n_p=Y_p \rho/m_p$ is the total number
density of protons, which make up a fraction $Y_p\approx 0.03$ of all
baryons.  The GL potential can be written as $U(\phi) = -a \phi^2 +(b/2)
\phi^4$.  

According to Ref.~\cite{at}, equation (\ref{du}) is valid for $T \ll T_c$,
which is the regime we are considering. The same equation  is also
valid~\cite{at} for $T\approx T_c$ if the characteristic frequency $\omega$
of time variations of the order parameter is greater than the energy gap,
$\omega > \Delta_0$.  This condition is automatically satisfied for nuclear
matter, where we find $\omega \sim 3$~MeV, while $\Delta_0 < 0.8$~MeV
(Fig.~1).    

Equation (\ref{du}) has a solution $\phi \propto \sin(\omega t)$
oscillating around the minimum of the effective potential 
$U(\phi)$. These oscillations are damped by the ``friction'' term in
equation~(\ref{du}) that contains $(\dot \phi)$.  The effect of damping
depends on the relative magnitudes of $8\varepsilon_{_F}/3 c \approx 7.4
T_c \approx 4.2 \Delta_0$ and the frequency of oscillations $\omega$. The
Fermi energy $\varepsilon_{_F}=p_{_F}^2/2m_* \approx 0.2$~to~20~MeV for
$\rho=10^{14}$ to $10^{15}$~g~cm$^{-3}$.  The oscillation frequency is 
determined by the effective mass $\omega = \sqrt{U''(\phi_0)}$ 
around the minimum $\phi_0 =\sqrt{a/b} $ of $U(\phi)$.  Requiring that
$U(\phi_0)= - \delta E/V$ from equation (\ref{dE}), one obtains
$\omega=\Delta_0\sqrt{2 m_*   p_{_F}}/\pi \phi_0  \approx 3$--$5$~MeV for
$\Delta_0=0.4$--0.8~MeV. Since $(8\varepsilon_{_F}/3 c)/\omega \approx
0.6$, neither the first, nor the second term in equation (\ref{du}) can be
neglected in our case. 

A coherent state of photons is described by the electromagnetic field
$A_\mu $. The Cooper pair of protons has charge $2e$.  
In the unitary gauge, in which the scalar field $\phi$ is real,  
the coupling of the gauge field to $\phi$ is  ${\cal L}_{int} = 
(2e)^2 \phi^2 (A_\mu)^2$.  

When $\phi$ oscillates around the minimum, the effective photon mass, 
$\sqrt{2 (2e)^2 \phi^2}$, is time-dependent.  This causes a parametric
resonance manifest in the appearance of exponentially growing solutions for
the gauge field $A_\mu$ interpreted as a copious non-thermal particle
production ({\it cf.} Ref.~\cite{cosmo}). We will assume a simple form for
the oscillations of the field $\f$ around the minimum, $\f = \f_{0}+\Phi 
\sin (\omega t)$, $\Phi \ll \phi_0 \approx 20$~MeV, and neglect the
back-reaction of photons on the motion of $\phi$.    
To see which modes are amplified, we consider a Fourier decomposition of
$A_\mu(x,t)$.  The equation of motion for a mode $A^{(k)}_\mu (t)$ with
the wavenumber $k$ is  

\beq
\frac{\dd^2}{\dd t^2} A^{(k)}_\mu  + [k^2 + 2 (2 e)^2 \f_0^2 + 
4 (2e)^2 \phi_0 \, \Phi \sin(\omega t)] A^{(k)}_\mu =0.
\label{Matheq}
\eeq 

The Matheiu equation (\ref{Matheq}) has exponentially growing solutions 
$A^{(k)} \propto \exp \{ \mu_k \omega t/2\}$ that describe a copious
production of the gamma-quanta with momenta in some spectral band. 
The stability  chart (see, {\it e.\,g.},  Ref. \cite{math}) is used to
determine the values of the momenta for which the resonant production takes
place.  The corresponding instability band is usually specified~\cite{math}
in terms of a parameter $a_k\equiv 4[(k^2+8 e^2 \f_0^2)/\omega^2]$.  If
the parameter $q \equiv 32 e^2 \phi_0 \Phi /\omega^2$ is small, only a
narrow band around $a_k= l^2$, where $l$ is some integer number, is in
resonance; the best-amplified mode has $k=\omega/2$ and $\mu_k=q/2$.  
If, however, $q>1$, the parametric resonance is broad.  A detailed
discussion of particle production through the parametric resonance can be
found in Ref.~\cite{cosmo}.   

For the proton condensate inside a neutron star, $\f_0 \sim 20$ MeV and
$\omega \sim 5$ MeV. If the amplitude of oscillations $\Phi$ is between 1 and
10 MeV, then $q \approx 2$--$20$, respectively.  This corresponds to a 
broad and, hence, very efficient resonance.  The exponentially growing
modes satisfy the resonance condition 

\beq
k^2 = \frac{1}{4} \omega^2  a_k- 8 e^2 \phi_0^2, 
\label{band} 
\eeq
where the resonant values of $a_k$ and the corresponding exponents $\mu_k$
can be read off the instability chart~\cite{math}.  For the ranges of the
parameters quoted above, a broad band is in resonance; the exponents $\mu_k$
can take relatively large values.  This signals a copious production of
gamma rays, whose occupation numbers increase exponentially already during
the first few oscillations of the field $\phi$.  As the amplitude of the
oscillations subsides, the system enters into a narrow-resonance regime, in
which only the modes around $k=\omega/2 \sim$ a few MeV are further
amplified.  Given the efficiency of the resonance and the absence of
thermal excitations at $T\ll T_c$ that could dissipate the energy stored in
the condensate, it is reasonable to assume that a substantial fraction of
the power $P$ in equation (\ref{power}) is transfered to photons.  

Because the electrons are highly degenerate, the gamma quanta cannot  
decay via the pair production $\gamma \rightarrow  e^+e^-$ unless the
final-state electrons have energy in excess of the Fermi momentum
$p_{_F} \stackrel{>}{_{\scriptstyle \sim}} 100$~MeV.  Therefore, a decay of
the MeV-energy gamma photons is forbidden by the Pauli exclusion principle
(applied to the final-state electrons).  Compton scattering off the
electrons and protons near the Fermi surface is kinematically suppressed
but is not forbidden.  It increases the temperature of the electrons and
makes the spectrum of the gamma rays 
somewhat more thermal by the time they reach the surface of the star.
Comptonization preserves the number of photons.  It can cause some
de-coherence effects that take place over the distances of order $10^{-9}$~cm,
much greater than the wavelength.  A coherent wave-packet passing through
neighbouring regions with oscillating scalar condensate may undergo further 
amplification.  The thin outer crust of a
neutron star is optically opaque.  It is not clear, however,  whether it
can withstand the photon pressure corresponding to the energy in equation
(\ref{power}) and prevent the gamma rays from leaving the star. 

The duration $\tau $ of resonant photo-production depends on how soon the
nuclear matter is heated to the temperature of order $T_c$.  It would seem  
that the detailed studies of coalescing neutron stars~\cite{cns} predict a
typical time scale of a few milliseconds for the acoustic shock waves to
heat the nuclear matter.  
However, for technical reasons, these simulations~\cite{cns} started with
non-zero initial temperatures to reduce the sensitivity of their
results to small variations of the internal energy at $T=0$.  In addition,
these analyses do not take into account the possible cooling effects of the
resonant gamma-emission which, as seen from equation (\ref{power}), can be
very significant and can delay the eruption of a thermal fireball.
Therefore, one can take $1$~ms as a very conservative lower
bound on the duration of the resonant production of gamma rays.  If a more
detailed analysis shows that $\tau$ can be as long as 1~s, the mechanism we
described can explain the origin of the observed GRB.  If $\tau \ll 1$~s,
one can look for a signature of resonant emission during the first
milliseconds of each GRB, assuming they originate from some events that
involve vibrating cold neutron stars.  

In summary, we have described a mechanism that can lead to a short burst of
gamma rays with power up to $10^{52\pm 1}$~erg~s$^{-1}$ from a 
jolted neutron star (for instance, in the event of a neutron star
collision).  The resonant emission of gamma rays is a result of a
non-thermal photo-production by the acoustically-powered coherent   
oscillations of the superconducting proton condensate.  The duration of the
pulse is uncertain and depends on the details of the heat transport.  
It is possible that the observed gamma-ray bursts, or at least
the signal during the short time interval in the beginning of some bursts,
may be explained by this mechanism. 

The author thanks A.~De~R\'ujula, R.~Kamien, V.~Kuzmin, G.~Segr\`e, and
M.~Shaposhnikov for helpful discussions.

\end{document}